\documentclass[pra,floatfix,showpacs,superscriptaddress,amsmath,twocolumn]{revtex4}
\usepackage{amssymb}
\usepackage{graphics}
\usepackage{epsfig}
\usepackage{graphicx}

\usepackage{dcolumn}
\usepackage{amsmath,amssymb}
\usepackage{bm}
\usepackage[dvips]{color}

\begin{document}

\definecolor{mygrey}{gray}{0.35}
\definecolor{mygreen}{rgb}{0.85,1,0.9}
\definecolor{myzard}{cmyk}{0,0,0.05,0}
\definecolor{mywhite}{rgb}{1,1,1}
\definecolor{myred}{rgb}{1,0,0}

\def\C{{\mathbb{C}}} \def\F{{\mathbb{F}}}
\def\N{{\mathbb{N}}} \def\Q{{\mathbb{Q}}}
\def\R{{\mathbb{R}}} \def\Z{{\mathbb{Z}}}

\def\cH{{\mathcal H}}

\def\bfsigma{\boldsymbol{\sigma}}
\def\bfmu{\boldsymbol{\mu}}

\def\dd{\mathord{\rm d}} \def\Det{\mathop{\rm Det}}
\def\dist{\mathop{\rm dist}} \def\ee{\mathord{\rm e}}
\def\End{\mathord{\rm End}} \def\ev{\mathord{\rm ev}}
\def\id{\mathord{\rm id}} \def\ii{\mathord{\rm i}}
\def\min{\mathord{\rm min}} \def\mod{\mathord{\rm mod}}
\def\prob{\mathord{\rm prob}} \def\tr{\mathop{\rm Tr}}
\def\half{\textstyle\frac{1}{2}} \def\third{\textstyle\frac{1}{3}}
\def\fourth{\textstyle\frac{1}{4}}

\def\vec#1{{\bf{#1}}} \def\vect#1{\vec{#1}}

\def\bra#1{\langle#1|} \def\ket#1{|#1\rangle}
\def\braket#1#2{\langle#1|#2\rangle}
\def\bracket#1#2{\langle#1,#2\rangle} \def\ve#1{\langle#1\rangle}

\def\lR{l^2_{\mathbb{R}}}
\def\RR{\mathbb{R}}
\def\E{\mathbf e}
\def\D{\boldsymbol \delta}
\def\S{{\cal S}}
\def\T{{\cal T}}
\def\dd{\delta}
\def\one{{\bf 1}}
\def\ss{\boldsymbol \sigma}

\newtheorem{theorem}{Theorem}
\newtheorem{lemma}{Lemma}
\newtheorem{definition}{Definition}

\title{Spin entanglement loss by local correlation transfer to the momentum}

\author{Lucas Lamata} \email{lamata@imaff.cfmac.csic.es}
\affiliation{Instituto de Matem\'aticas y F\'{\i}sica Fundamental,
CSIC, Serrano 113-bis, 28006 Madrid, Spain}
\author{Juan Le\'on}\email{leon@imaff.cfmac.csic.es}
\affiliation{Instituto de Matem\'aticas y F\'{\i}sica Fundamental,
CSIC, Serrano 113-bis, 28006 Madrid, Spain}
\author{David Salgado}\email{david.salgado@uam.es}
\affiliation{Dpto. F\'{\i}sica Te\'orica, Universidad Aut\'onoma de
Madrid, 28049 Cantoblanco, Madrid, Spain}


\begin{abstract}
We show the decrease of spin-spin entanglement between two
$s=\frac{1}{2}$ fermions or two photons due to local transfer of
correlations from the spin to the momentum degree of freedom of one
of the two particles. We explicitly show how this phenomenon
operates in the case where one of the two fermions (photons) passes
through a local homogeneous magnetic field (optically-active
medium), losing its spin correlations with the other particle.
\end{abstract}

\maketitle
\section{INTRODUCTION}
Bipartite and multipartite entanglement is considered a basic
resource in most applications of quantum information, communication,
and technology~ (see for instance Refs. \cite{nielsen,GM02}).
Entanglement is fragile, and it is well-known that in some cases
interactions with an environment external to the entangled systems
may decrease the quantum correlations, degrading this valuable
resource \cite{Ye02,Ye03,Ye04,DH04,NS05,CMPB05,FMDZ05}. Momentum
acts as a very special environment which every particle possesses
and cannot get rid of. Consider for instance a bipartite system
which initially is spin-entangled and with the momentum
distributions factorized. It will decrease its spin-spin
correlations provided any of the two particles entangles its spin
with its momentum. This simple idea was studied in the natural
framework of special relativity, where changing the reference frame
induces Wigner rotations that entangle each spin with its momentum
\cite{PST02,C05,AM02,GA02,PS03,C03,GBA03,LMDS05}. However, this is
just a kinematical, frame-dependent effect only, and not a real
dynamical interaction. On the other hand, this type of reasoning is
also related to which-path detection \cite{SZ97}. In addition, an
experiment observing photon polarization disentanglement by
correlation transfer, in this case to the photon's position, was
performed \cite{FMKS01}. Can local interactions entangling spin with
momentum produce the loss of non-local spin-spin entanglement? To
our knowledge this question has not been explored in the literature.
More remarkably, any particle owns a certain momentum distribution
acting as an intrinsic environment, which can never be eliminated by
improving the experimental conditions. But how does this fact affect
the spin-spin correlations? This is the question we want to analyze
in this paper.

In Sec. \ref{lbsetc2} we consider a bipartite system, composed by
two $s=\frac{1}{2}$ fermions or two photons, which are initially in
a Bell spin state $|\Psi^{-}\rangle$. We use a formalism
\cite{LMDS05} that shows the decrease of spin entanglement whenever
an interaction locally entangling spin with momentum takes place. We
obtain the negativity $N$ \cite{VW02} in terms of an integral
depending on the spin rotation angle conditional to the momentum. In
Sec. \ref{lbsetc3} we analyze this physical phenomenon in two
specific situations: (i) Two spin-$\frac{1}{2}$ fermions in a
$|\Psi^{-}\rangle$ Bell state, with Gaussian momentum distributions,
that fly apart while one of them passes through a local magnetic
field. Their spin entanglement decreases as a consequence of the
transfer of correlations to the momentum of the latter fermion. And
(ii) Two photons in a polarization $|\Psi^{-}\rangle$ Bell state,
with Gaussian momentum distributions, which separate while one of
them traverses an optically-active medium. This medium will entangle
the polarization with the momentum and thus decrease the
polarization entanglement. This is an unavoidable source of
decoherence. In Sec. \ref{lbsetc4} we show that the apparent purely
quantum communication resulting from these procedures is not
possible. Classical communication has to be exchanged for the
protocol to operate. The paper ends with our conclusions in Sec.
\ref{lbsetc5}.
\section{SPIN ENTANGLEMENT LOSS BY CORRELATION TRANSFER TO
THE MOMENTUM\label{lbsetc2}}

We consider a maximally spin-entangled state for two $s=\frac{1}{2}$
fermions $A$ and $B$, or two photons $A$ and $B$. The case we
analyze is that in which the two particles are far apart already.
This avoids dealing with symmetrization issues. Indeed, our state is
the maximally entangled one for two $s=\frac{1}{2}$ spins or
polarizations, containing 1 ebit.

\begin{eqnarray}
| \Psi^{-}_{\vec{p}} \rangle \!\!\! &  := & \!\!\!
\frac{1}{\sqrt{2}}\lbrack \Psi_\uparrow^{(\rm a)} (\vec{p_a})
\Psi_\downarrow^{(\rm b)} (\vec{p_b}) -\Psi_\downarrow^{(\rm a)}
(\vec{p_a}) \Psi_\uparrow^{(\rm b)} (\vec{p_b})\rbrack ,  \nonumber \\
 \label{deco1}
\end{eqnarray}
where $\vec{p_a}$ and $\vec{p_b}$ are the corresponding momentum
vectors of particles $A$ and $B$, and
\begin{eqnarray}
\Psi_\uparrow (\vec{p}) &:=&  {\cal M} ( \vec{p} ) |\!\!\uparrow
\rangle = \left(
\begin{array}{cccc}
{\cal M} ( \vec{p} ) \\
0 \\
\end{array}
\right), \nonumber \\
\Psi_\downarrow (\vec{p}) &:=&  {\cal M} ( \vec{p} ) |\!\!\downarrow
\rangle = \left(
\begin{array}{cccc}
0 \\
{\cal M} ( \vec{p} ) \\
\end{array}
\right), \label{deco2}
\end{eqnarray}
with bimodal momentum distribution ${\cal M} ( \vec{p} ):=
\frac{1}{\sqrt{2}}(\delta_{\vec{p}\vec{p_1}}+\delta_{\vec{p}\vec{p_2}})$.
$\vec{p}_1$ and $\vec{p}_2$ are the two momentum values considered
associated to particle $A$, $\vec{p}_1^{(\rm a)}$, $\vec{p}_2^{(\rm
a)}$, on the one hand, and particle $B$, $\vec{p}_1^{(\rm b)}$,
$\vec{p}_2^{(\rm b)}$, on the other hand. We consider for the time
being this kind of distribution for illustrative purposes. At the
end of this section we generalize our results to arbitrary momentum
distributions of the two particles. $|\!\!\uparrow \rangle$ and
$|\!\!\downarrow \rangle$ represent either spin vectors pointing up
and down along the $z$-axis, in the fermionic case, or right-handed
and left-handed circular polarizations, in the photonic case. If we
trace out the momentum degrees of freedom in Eq.~(\ref{deco1}), we
obtain the usual spin Bell state, $ | \Psi^{-} \rangle$.

We consider a local interaction which entangles the spin of each
particle with its momentum through a real unitary (orthogonal)
transformation $U$. We choose a real transformation for the sake of
simplicity and in order to obtain fully analytical results. The
generalization for inclusion of complex phases is straightforward
but adds nothing of relevance in this section. We will take it fully
into account in Sec. \ref{lbsetc3}.

\begin{figure}
\begin{center}
\includegraphics[width=7cm]{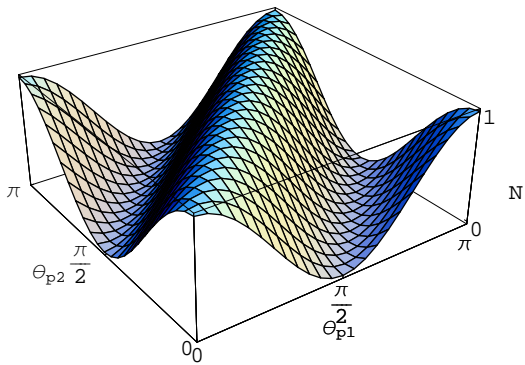}
\end{center}
\caption{(Color online) Negativity $N$ in Eq. (\ref{deco7}) as a
function of $\theta_{\vec{p_1}}$ and
$\theta_{\vec{p_2}}$.\label{grafdeco}}
\end{figure}

 Each state vector in Eq.~(\ref{deco2}) transforms as
\begin{eqnarray}
&&\Psi_\uparrow(\vec{p})= \left(
\begin{array}{cccc}
{\cal M} ( \vec{p} ) \\
0 \\
\end{array}
\right) \rightarrow \nonumber\\&& U[\Psi_\uparrow(\vec{p})]=\left(
\begin{array}{cccc}
\cos \theta_{\vec{p_1}} \\
\sin \theta_{\vec{p_1}} \\
\end{array}
\right)
\frac{\delta_{\vec{p}\vec{p_1}}}{\sqrt{2}}+\left(\begin{array}{cccc}
\cos \theta_{\vec{p_2}} \\
\sin \theta_{\vec{p_2}} \\
\end{array}
\right)\frac{\delta_{\vec{p}\vec{p_2}}}{\sqrt{2}},
\nonumber \\
&& \Psi_\downarrow(\vec{p})=\left(
\begin{array}{cccc}
0 \\
{\cal M} ( \vec{p} ) \\
\end{array}
\right) \rightarrow\nonumber\\&& U[\Psi_\downarrow(\vec{p})]=\left(
\begin{array}{cccc}
-\sin \theta_{\vec{p_1}} \\
\,\,\,\,\, \cos \theta_{\vec{p_1}} \\
\end{array}
\right)
\frac{\delta_{\vec{p}\vec{p_1}}}{\sqrt{2}}+\left(\begin{array}{cccc}
-\sin \theta_{\vec{p_2}} \\
\,\,\,\,\, \cos \theta_{\vec{p_2}} \\
\end{array}
\right)  \frac{\delta_{\vec{p}\vec{p_2}}}{\sqrt{2}} ,\nonumber\\
\label{deco3}
\end{eqnarray}
where $\theta_{\vec{p_1}}$ and $\theta_{\vec{p_2}}$ produce a
spin-momentum entangled state whenever
$\theta_{\vec{p_1}}\neq\theta_{\vec{p_2}}$. The effect of this local
interaction is that a part of the non-local spin-spin entanglement
is transferred to the spin-momentum one, and the degree of
entanglement of the spins decreases. To show this, we consider the
state (\ref{deco1}) evolved with the transformation $U$, and trace
out the momenta.
\begin{eqnarray}
& \mathrm{Tr}_{\rm \vec{p_a}, \vec{p_b}}&  (U| \Psi^{-}_{\vec{p}}
\rangle \langle \Psi^{-}_{\vec{p}} |  U^{\dag})  \nonumber\\ & =
&\frac{1}{2}\;\sum_{s,s'}ss' \mathrm{Tr}_{\rm \vec{p_a}}(U^{(\rm
a)}[\Psi^{(\rm a)}_s( \vec{p_a} )]\{U^{(\rm a)}[\Psi^{(\rm a)}_{s'}(
\vec{p_a})]\}^{\dag}) \nonumber\\&&\otimes \mathrm{Tr}_{\rm
\vec{p_b}}(U^{(\rm b)}[\Psi^{(\rm b)}_{-s}( \vec{p_b} )]\{U^{(\rm
b)}[\Psi^{(\rm b)}_{-s'}( \vec{p_b})]\}^{\dag}),\label{deco4}
\end{eqnarray}
where $ss':=\delta_{s,s'}-\delta_{s,-s'}$.
 It can be appreciated in Eq. (\ref{deco4}) that the expression is decomposable in sum of tensor products of 2$\times$2 spin blocks,
 each corresponding to each particle. We compute now the different
 blocks, corresponding to the four possible tensor
 products of the states (\ref{deco3})
\begin{eqnarray}
\mathrm{Tr}_{\rm \vec{p}}[U\Psi_\uparrow(U\Psi_\uparrow)^{\dag}] & =
& \frac{1}{2} \left(\begin{array}{cc}c_1^2+c_2^2 & c_1 s_1+c_2
s_2\\c_1 s_1+c_2 s_2 & s_1^2+s_2^2\end{array}\right),\nonumber
\\
\mathrm{Tr}_{\rm \vec{p}}[U\Psi_\uparrow(U\Psi_\downarrow)^{\dag}]&
= &\frac{1}{2}\left(\begin{array}{cc}-c_1 s_1-c_2 s_2 &
c_1^2+c_2^2\\-s_1^2-s_2^2 & c_1 s_1+c_2
s_2\end{array}\right),\nonumber\\
\mathrm{Tr}_{\rm \vec{p}}[U\Psi_\downarrow(U\Psi_\uparrow)^{\dag}]&=
&\frac{1}{2}\left(\begin{array}{cc}-c_1 s_1-c_2 s_2 &
-s_1^2-s_2^2\\c_1^2+c_2^2 &c_1 s_1+c_2
s_2\end{array}\right),\nonumber\\
\mathrm{Tr}_{\rm
\vec{p}}[U\Psi_\downarrow(U\Psi_\downarrow)^{\dag}]&=&\frac{1}{2}
\left(\begin{array}{cc}s_1^2+s_2^2 & -c_1 s_1-c_2 s_2\\-c_1 s_1-c_2
s_2 & c_1^2+c_2^2\end{array}\right),\nonumber
\end{eqnarray}
where $c_i  :=  \cos(\theta_{\vec{p_i}})$ and $s_i  :=
\sin(\theta_{\vec{p_i}})$. This way, it is possible to compute the
effects of the local interaction $U$ in the state $|
\Psi^{-}_{\vec{p}} \rangle$ after tracing out the momentum. We
choose equal interaction angles for the two particles,
$\theta_{\vec{p_i}}^{(\rm a)}=\theta_{\vec{p_i}}^{(\rm b)}$, as a
natural simplification. The resulting bipartite spin state is
\begin{equation}
 \left( \!\!\!
\begin{array}{cccc}
\frac{1}{4}s_{12}^2& 0 & 0 &
\frac{1}{4}s_{12}^2\\
0 & \frac{1}{4}(1+ c_{12}^2)&
-\frac{1}{4}(1+ c_{12}^2)& 0 \\
0 & -\frac{1}{4}(1+ c_{12}^2) & \frac{1}{4}(1+ c_{12}^2) & 0 \\
\frac{1}{4}s_{12}^2 & 0 & 0 &
\frac{1}{4}s_{12}^2 \\
\end{array}
\!\!\! \right) , \\ \label{deco6}
\end{equation}
where $s_{12}:=\sin(\theta_{\vec{p_1}}-\theta_{\vec{p_2}})$ and
$c_{12}:=\cos(\theta_{\vec{p_1}}-\theta_{\vec{p_2}})$. The
entanglement monotone we will use is the negativity \cite{VW02},
defined as $N:=\max\{0,-2\lambda_{\rm min}\}$, where $\lambda_{\rm
min}$ is the smallest eigenvalue of the partial transpose (PT)
matrix of Eq. (\ref{deco6}). It is very easily computable, and is
found to be
\begin{equation}
 N=\cos^2(\theta_{\vec{p_1}}-\theta_{\vec{p_2}}).\label{deco7}
\end{equation}
From this expression it can be appreciated that for
$\theta_{\vec{p_1}}=\theta_{\vec{p_2}}$ the entanglement remains
maximal (1 ebit), and for $\theta_{\vec{p_1}}$ separating from
$\theta_{\vec{p_2}}$ the entanglement decreases, until
$\theta_{\vec{p_1}}-\theta_{\vec{p_2}}=\pi/2$, where it vanishes and
the state becomes separable. We plot this behavior in Figure
\ref{grafdeco}, showing the negativity $N$ in Eq. (\ref{deco7}) as a
function of $\theta_{\vec{p_1}}$ and $\theta_{\vec{p_2}}$. Every
local interaction producing spin-momentum entanglement will in
general diminish the initial maximal spin-spin entanglement of the
two particles, thus degrading this resource. This result is valid
either for $s=\frac{1}{2}$ fermions or photons, as they both have a
two-dimensional internal Hilbert space.
The generalization of Eq. (\ref{deco7}) to a uniform distribution
with $n$ different momenta is straightforward, and the spectrum of
the corresponding PT matrix is
\begin{equation}
\sigma_{\rm PT}=\left\{\frac{1}{2},\frac{1}{2},\pm
\left[\frac{1}{2}-\frac{1}{n^2}\sum_{i,j=1}^n\cos^2(\theta_{\vec{p_i}}-\theta_{\vec{p_j}})\right]\right\},\label{deco8}
\end{equation}
with a resulting negativity
\begin{equation}
N=\left|1-\frac{2}{n^2}\sum_{i,j=1}^n\cos^2(\theta_{\vec{p_i}}-\theta_{\vec{p_j}})\right|.\label{deco9}
\end{equation}
We take now the continuous limit, for an arbitrary momentum
distribution $\tilde{\psi}(\vec{p})$ for each particle. We suppose
for the sake of simplicity $|\tilde{\psi}^{(\rm
a)}(\vec{p})|=|\tilde{\psi}^{(\rm b)}(\vec{p})|$, although the
spatial distributions do not overlap, as the two particles are far
away. Accordingly, $N$ will be
\begin{equation}
N=\left|1-2\int d^3\vec{p}\int
d^3\vec{p'}|\tilde{\psi}(\vec{p})|^2|\tilde{\psi}(\vec{p'})|^2
\cos^2(\theta_{\vec{p}}-\theta_{\vec{p'}})\right|.\label{deco10}
\end{equation}
Notice that this expression involves an integration over the
momentum variables $\vec{p}$ and $\vec{p'}$, associated to the same
particle (not to each of them). We point out that, according to
(\ref{deco10}), in the case where momentum does not entangle with
spin (i.e. whenever $\theta_{\vec{p}}$ is a constant), then $N=1$
and thus the spin-spin entanglement remains maximal. Otherwise, the
spin-spin entanglement would decrease due to the transfer of
correlations to the spin-momentum part.

The loss of spin entanglement under a spin-momentum entangling
transformation can take place in a variety of possible situations.
Wigner rotations that appear under relativistic change of reference
frame entangle each spin with its momentum producing loss of
spin-spin entanglement
\cite{PST02,C05,AM02,GA02,PS03,C03,GBA03,LMDS05}. This is just a
kinematical-relativistic effect, not due to a dynamical interaction.
In the rest of the paper we focus on two relevant examples of these
interactions, taking fully into account the complex phases: a local
homogeneous magnetic field, for the two-fermion case, and a local
optically-active medium, for the two-photon case.

\section{APPLICATIONS\label{lbsetc3}}
\subsection{Two fermions and a local magnetic field}

In this section we analyze a bipartite system, composed by two
$s=\frac{1}{2}$ neutral fermions $A$ and $B$, which are initially
far apart and in a Bell spin state $|\Psi^{-}\rangle$, with
factorized Gaussian momentum distributions. We consider that one of
them traverses a region where a finite, homogeneous magnetic field
exists. As a result, it will transfer part of its spin correlations
to the momentum.

The initial spin-entangled state for the two fermions $A$ and $B$ is

\begin{eqnarray}
| \Psi^{-}_{\vec{p}} \rangle \!\!\! &  := & \!\!\!
\frac{1}{\sqrt{2}}\lbrack \Psi_\uparrow^{(\rm a)} (\vec{p_a})
\Psi_\downarrow^{(\rm b)} (\vec{p_b}) -\Psi_\downarrow^{(\rm a)}
(\vec{p_a}) \Psi_\uparrow^{(\rm b)} (\vec{p_b})\rbrack ,  \nonumber \\
 \label{v2deco1}
\end{eqnarray}
where $\vec{p_a}$ and $\vec{p_b}$ are the corresponding momentum
vectors of particles $A$ and $B$, and
\begin{eqnarray}
\Psi_\uparrow (\vec{p}) &:=&  {\cal G} ( \vec{p} ) |\!\!\uparrow
\rangle = \left(
\begin{array}{cccc}
{\cal G} ( \vec{p} ) \\
0 \\
\end{array}
\right), \nonumber \\
\Psi_\downarrow (\vec{p}) &:=&  {\cal G} ( \vec{p} ) |\!\!\downarrow
\rangle = \left(
\begin{array}{cccc}
0 \\
{\cal G} ( \vec{p} ) \\
\end{array}
\right), \label{v2deco2}
\end{eqnarray}
with Gaussian momentum distribution ${\cal G} ( \vec{p} ):=
\pi^{-3/4}\sigma^{-3/2} \exp [ -  (\vec{p}-\vec{p}_0)^2/2 \sigma^2
]$. In Eqs. (\ref{v2deco2}) we are not indicating explicitly the
particle index. In the center of mass frame, $\vec{p}^{(\rm
b)}_0=-\vec{p}^{(\rm a)}_0$, and we consider that the two particles
are flying apart from each other. $|\!\!\uparrow \rangle$ and
$|\!\!\downarrow \rangle$ represent spin vectors pointing up and
down along the $z$-axis, respectively. If we trace out momentum
degrees of freedom in Eq.~(\ref{v2deco1}), we obtain the usual spin
Bell state, $ | \Psi^{-} \rangle$.

\begin{figure}
\begin{center}
\includegraphics[width=5.5cm]{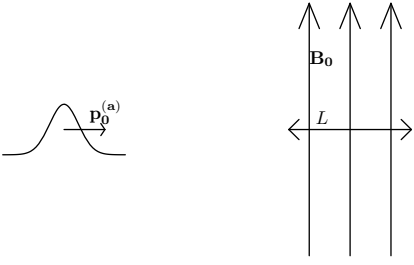}
\end{center}
\caption{Sketch of the two-fermion case explained in the text.
Fermion $A$ traverses a constant magnetic field $\vec{B}_0$ located
in region $\cal{D}$ with a width $L$ along the direction of
$\vec{p}^{(\rm a)}_0$. \label{v2grafdeco1}}
\end{figure}

Suppose a local interaction which entangles the spin of fermion $A$
with its momentum through a unitary transformation $U$. In this case
we choose a magnetic field $\vec{B}_0$ which is constant on a
bounded region $\cal{D}$ of length $L$, along the direction of
$\vec{p}^{(\rm a)}_0$, vanishes outside $\cal{D}$, and extends
infinitely with a constant value along the other two orthogonal
directions, as shown in Figure \ref{v2grafdeco1}. We take
$\vec{B}_0$ along the direction orthogonal to $\vec{p}^{(\rm a)}_0$,
so it is divergenceless, $\vec{\nabla}\cdot \vec{B}_0=0$, and we
quantize the spin along $\vec{B}_0$ so that
$\vec{\sigma}\cdot\vec{B}_0=s B_0$, with $s$ the corresponding spin
component. Due to the rotational invariance of the spin singlet,
this choice is completely general. In momentum space, the problem
reduces to one dimension, the one associated to the direction of
$\vec{p}^{(\rm a)}_0$. We will denote from now on $p$ to the
corresponding momentum coordinate.

The system Hamiltonian can be written as
\begin{equation}
H\,=\, \frac{\vec{p}^2}{2 m}\,+\, \gamma \,(\vec{B}_0\cdot\vec{S})
\, \theta(x)\, \theta(L-x),\label{v3deco1}
\end{equation}
where $\gamma$ is the magnetic moment of our neutral particle.
Accordingly, we get
\begin{eqnarray}
\dot{\vec{p}}\,=\,i[H,\vec{p}]&=&\left(-\gamma\, (\vec{B}_0\cdot \vec{S})\, [\delta(x)-\delta(L-x)],0,0\right),\label{v3deco2}\\
\dot{\vec{S}}\,=\, i[H,\vec{S}]&=& -\gamma\,(\vec{B_0}\wedge\vec{S})\, \theta(x)\,\theta(L-x),\nonumber\\
\ddot{\vec{S}}\,=\, i[H,\dot{\vec{S}}]&=&
-\gamma\,\left[(\vec{B_0}\cdot\vec{S})\,\vec{B_0}- \vec{B_0}^2
\vec{S}\right]\theta(x)\,\theta(L-x)\nonumber\\&&-\,i
\gamma\,(\vec{B_0}\wedge\vec{S}) \left(\frac{\vec{p}}{m}
[\vec{p},(\delta(x)-\delta(L-x))]\right).\nonumber
\end{eqnarray}
From the first of the above equations we obtain
$\frac{\partial}{\partial
x}(p^2/2m)=-\gamma\,(\vec{B_0}\cdot\vec{S})
\left[\delta(x)-\delta(L-x)\right]$ like using matching conditions
at $x=0,L$. The second and third equations give the spin evolution.
By inspection we see that i) the spin remains parallel to
$\vec{B_0}$ if it was initially so and, ii) the spin is constant in
this case $\dot{\vec{S}}\,=\,\ddot{\vec{S}}\,=\,0$. Hence, in spite
of choosing a case where the spin is conserved, its entanglement
with the momentum decreases the spin correlations with the idle
fermion.

 The effect of the magnetic field on particle $A$ can be seen in its state. Behind the region
 $\cal{D}$, the resulting state vector as transformed from the one in Eq.~(\ref{v2deco2})
 is
\begin{eqnarray}
&&\Psi_\uparrow(\vec{p})\rightarrow
U[\Psi_\uparrow(\vec{p})]={\cal{T}}_{\uparrow}(p)\left(
\begin{array}{cccc}
{\cal G} ( \vec{p} ) \\
0 \\
\end{array}
\right),
\nonumber \\
&& \Psi_\downarrow(\vec{p})\rightarrow
U[\Psi_\downarrow(\vec{p})]={\cal{T}}_{\downarrow}(p)\left(
\begin{array}{cccc}
0 \\
{\cal G} ( \vec{p} ) \\
\end{array}\right) ,\nonumber\\
\label{v2deco3}
\end{eqnarray}
where ${\cal{T}}_{\uparrow}(p)$ (${\cal{T}}_{\downarrow}(p)$) is the
transmission coefficient associated to the mesa (well) potential
induced by $\vec{B}_0$, for initial spin up (down). It is given by
\begin{equation}
{\cal{T}}_s(p):=\frac{2 p p_s e^{-i p L}}{2 p p_s\cos(p_s
L)-i(p^2+p_s^2)\sin(p_s L)}, \label{TransCoef}
\end{equation}
where $p_s(p,B_0):=\sqrt{p^2-2 s m \gamma B_0}$, as given by
(\ref{v3deco2}), $B_0:=|\vec{B_0}|$ and
$s=\frac{1}{2}(-\frac{1}{2})$ for spin $\uparrow(\downarrow)$. As
expected for $B_0=0$,
${\cal{T}}_{\uparrow}(p)={\cal{T}}_{\downarrow}(p)=1$. The initial
state is preserved so no spin-momentum correlations are generated.
In general, for $B_0\neq 0$,
${\cal{T}}_{\uparrow}(p)\neq{\cal{T}}_{\downarrow}(p)$, producing
spin-momentum entanglement. We are considering here just
transmission through the region $\cal{D}$, without taking into
account the reflection of the wave packets. We suppose all the
measurements will take place beyond $\cal{D}$ so we may normalize
the final transmitted state to 1. Finally, the net effect of this
local interaction is the reshuffling of spin-momentum correlations
in the state of the active fermion. Accordingly, the degree of
spin-spin entanglement between both particles decreases. As was done
in Eq. (\ref{deco4}), we evolve the state (\ref{v2deco1}) with the
transformation $U$ and trace out the momenta
\begin{eqnarray}
& \mathrm{Tr}_{\rm \vec{p_a}, \vec{p_b}} & (U| \Psi^{-}_{\vec{p}}
\rangle \langle \Psi^{-}_{\vec{p}} |  U^{\dag})\nonumber
\\  & = & \!\!\!\!\!\!\frac{1}{2}\sum_{s,s'}ss'\mathrm{Tr}_{\rm
\vec{p_a}}(U[\Psi^{(\rm a)}_s( \vec{p_a} )]\{U[\Psi^{(\rm a)}_{s'}(
\vec{p_a})]\}^{\dag}) \nonumber\\&&\otimes \mathrm{Tr}_{\rm
\vec{p_b}}(\Psi^{(\rm b)}_{-s}( \vec{p_b} )\{\Psi^{(\rm b)}_{-s'}(
\vec{p_b})\}^{\dag}),\label{v2deco4}
\end{eqnarray}
where $ss':=\delta_{s,s'}-\delta_{s,-s'}$. The traces corresponding
to particle $B$ give just the initial spin states,
$|\!\!\!\downarrow\rangle\langle\downarrow\!\!\!|$,
$|\!\!\!\uparrow\rangle\langle\uparrow\!\!\!|$,
$|\!\!\!\downarrow\rangle\langle\uparrow\!\!\!|$,
$|\!\!\!\uparrow\rangle\langle\downarrow\!\!\!|$, because $U$ is
just the identity for $B$. The resulting, properly normalized
spin-spin state is
\begin{equation}
 \left( \!\!\!
\begin{array}{cccc}
0& 0 & 0 &
0\\
0 & I_{\uparrow\uparrow}&
-I_{\uparrow\downarrow}& 0 \\
0 & -I_{\downarrow\uparrow} & I_{\downarrow\downarrow} & 0 \\
 0 & 0 & 0 & 0
 \\
\end{array}
\!\!\! \right) , \\ \label{v2deco6}
\end{equation}
where
\begin{eqnarray}
I_{ss'}:=\int d^3 \vec{p} |{\cal{G}}(\vec{p})|^2
\frac{{\cal{T}}_s(p){\cal{T}}^*_{s'}(p)}{|{\cal{T}}_{\uparrow}(p)|^2+|{\cal{T}}_{\downarrow}(p)|^2}.\label{v2deco7}
\end{eqnarray}
The negativity for this state is found to be
\begin{equation}
N=2|I_{\uparrow\downarrow}|.\label{v2deco8}
\end{equation}
 This
expression for $N$ is rather illuminating and its behavior easy to
understand. For the initial state (\ref{v2deco1}) $N=1$ (1 initial
ebit), and, as long as the magnetic field is increased,
${\cal{T}}_{\uparrow}(p)$ and ${\cal{T}}_{\downarrow}(p)$ become
more different, making the term $I_{\uparrow\downarrow}$ smaller,
and diminishing $N$.
\begin{figure}
\begin{center}
\includegraphics[width=7cm]{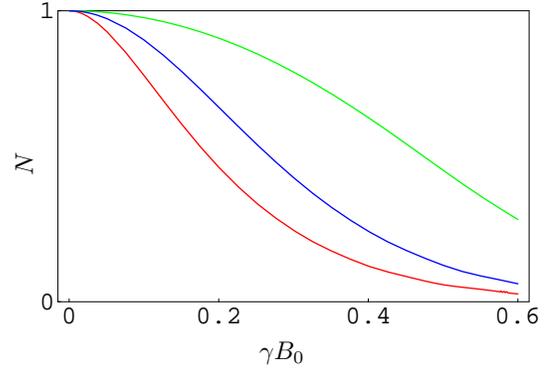}
\end{center}
\caption{(Color online) Negativity $N$ in Eq. (\ref{v2deco8}) as a
function of $\gamma B_0$ for $m=100$, $p^{(\rm a)}_0=10$, $L=3$, and
$\sigma^{(\rm a)}=1$, 2, and 3. The higher curves correspond to the
thinner $\sigma$'s. All quantities are given in $0.1 p_0^{(\rm a)}$
units. \label{v2grafdeco2}}
\end{figure}
\begin{figure}
\begin{center}
\includegraphics[width=7cm]{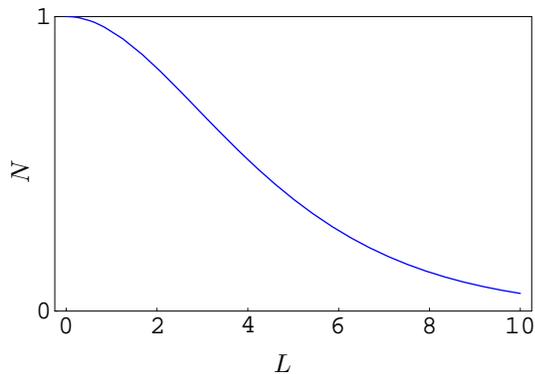}
\end{center}
\caption{(Color online) Negativity $N$ in Eq. (\ref{v2deco8}) as a
function of $L$ for $m=100$, $p^{(\rm a)}_0=10$, $\gamma B_0=0.2$,
and $\sigma^{(\rm a)}=2$. All quantities are given in $0.1 p_0^{(\rm
a)}$ units. \label{v2grafdeco3}}
\end{figure}
On the other hand, the wider $\sigma^{(\rm a)}$  the more
destructive interference between ${\cal{T}}_{\uparrow}(p)$ and
${\cal{T}}_{\downarrow}(p)$ will occur, reducing $N$. We plot in
Figure \ref{v2grafdeco2} the negativity $N$ as a function of $\gamma
B_0$, with $B_0:=|\vec{B}_0|$, for $m=100$, $p^{(\rm
a)}_0:=|\vec{p}^{(\rm a)}_0|=10$, $L=3$, and $\sigma^{(\rm a)}=1$,
2, and 3. All quantities are given in $0.1 p_0^{(\rm a)}$ units. The
entanglement goes to zero with increasing $B_0$, and the wider
$\sigma^{(\rm a)}$, the lesser the entanglement. A similar behaviour
arises from the cumulative effect of the barrier; the larger $L$,
the smaller the entanglement. We show in Figure \ref{v2grafdeco3}
this behavior, plotting $N$ as a function of $L$ for $m=100$,
$p^{(\rm a)}_0=10$, $\gamma B_0=0.2$, and $\sigma^{(\rm a)}=2$.
\subsection{Two photons and an optically-active medium}

In this section we analyze a bipartite system, composed by two
photons $A$ and $B$, which are far apart in a polarization Bell
state $|\Psi^{-}\rangle$ with factorized Gaussian momentum
distributions. We consider that the photon $A$ traverses a region
where a finite, optically-active medium, exists. As a result, part
of its spin correlations will be transferred to the momentum.

Basically the mathematical formalism used for the two-fermion case
is also valid here, with $\uparrow$ and $\downarrow$ indicating
right-hand and left-hand circular polarizations, which we will
denote by R and L indices. The transmission coefficient in the WKB
approximation, at lowest order, takes now the form of a complex
phase, depending on the polarization
\begin{equation}
{\cal{T}}_s(w):=\exp[i w n_s(w) L],\;\;\;s=R,L,\label{v2deco8ph}
\end{equation}
where the refraction indices are
\begin{equation}
n_{\rm R,L}(w):=\sqrt{1+\chi_{11}\pm\chi_{12}},\label{v2deco9ph}
\end{equation}
and $\chi_{11}$, $\chi_{12}$ are two of the matrix elements of the
susceptibility $\chi$,
\begin{equation}
\chi:=\left(\begin{array}{ccc}\chi_{11} & i\chi_{12} & 0\\-i
\chi_{12} & \chi_{11} & 0\\0 & 0 &
\chi_{33}\end{array}\right).\label{v2deco10ph}
\end{equation}
$\chi$ is produced, for example, by an isotropic dielectric placed
in a magnetic field $\vec{B}_0$ directed along $z$, which is also
the direction of photon propagation. $L$ is the dielectric length
that the photon traverses. $\chi_{11}$ and $\chi_{12}$ are
\begin{eqnarray}
&&\chi_{11}(w):=\frac{{\cal{N}}
e^2}{m\epsilon_0}\left[\frac{w_0^2-w^2}{(w_0^2-w^2)^2+w^2w_c^2}\right],\nonumber\\
&&\chi_{12}(w):=\frac{{\cal{N}}
e^2}{m\epsilon_0}\left[\frac{ww_c}{(w_0^2-w^2)^2+w^2w_c^2}\right],
\end{eqnarray}
where the cyclotron frequency $w_c:=e|\vec{B}_0|/m$, $e$ is the
electron charge, $m$ its mass, $w_0$ the resonance frequency of the
optically-active medium, ${\cal{N}}$ the number of electrons per
unit volume, and $\epsilon_0$ the vacuum electric permittivity.

 The next order correction would include factors
$\sqrt{n_{\rm R,L}}$ in the denominators of the transmission
coefficients. However, the approximation that considers these
factors coincides exactly with the lowest order one when taking into
account just linear terms in $\vec{B}_0$. We will consider the
realistic case in which $w_c$ is small as compared to the photon
average energy. Thus we will work from the very beginning just with
the transmission coefficients (\ref{v2deco8ph}).

The negativity $N$, obtained for this case in analogy with the
two-fermion case, is
\begin{equation}
N\simeq\frac{1}{\sqrt{\pi}\sigma}\left|\int_0^\infty dw
e^{-(w-p_0)^2/\sigma^2}e^{i \tilde{B}
L\frac{w^2}{(w^2-w_0^2)^2}}\right|,\label{v2deco11ph}
\end{equation}
where $p_0$ is the average momentum of photon $A$, $\sigma\ll p_0$
its momentum width, and $\tilde{B}:={\cal
N}e^3|\vec{B}_0|/(m^2\epsilon_0)$.

\begin{figure}
\begin{center}
\includegraphics[width=7cm]{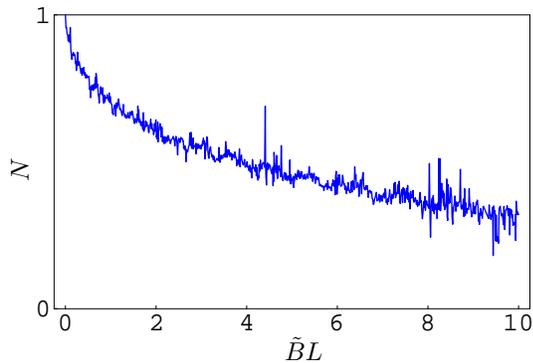}
\end{center}
\caption{(Color online) Negativity $N$ in Eq. (\ref{v2deco11ph}) as
a function of $\tilde{B}L$ for $p_0=10$, $\sigma=2$, and $w_0=10$.
All quantities are given in $0.1 p_0$ units.
\label{v2grafdecophoton1}}
\end{figure}

\begin{figure}
\begin{center}
\includegraphics[width=7cm]{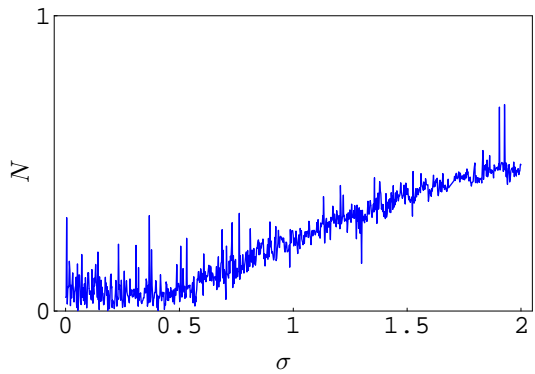}
\end{center}
\caption{(Color online) Negativity $N$ in Eq. (\ref{v2deco11ph}) as
a function of $\sigma$ for $p_0=10$, $\tilde{B}L=4$, and $w_0=10$.
All quantities are given in $0.1 p_0$ units.
\label{v2grafdecophoton2}}
\end{figure}

\begin{figure}[h]
\begin{center}
\includegraphics[width=7cm]{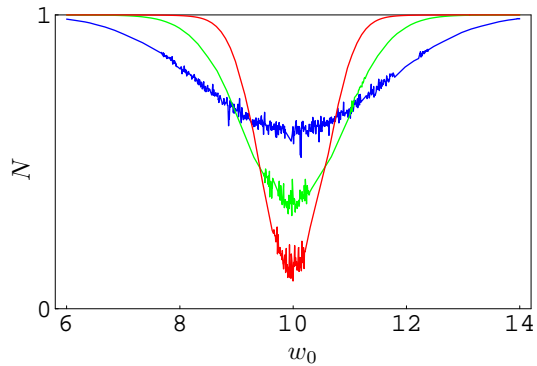}
\end{center}
\caption{(Color online) Negativity $N$ in Eq. (\ref{v2deco11ph}) as
a function of $w_0$ for $p_0=10$, $\tilde{B}L=2$, and
$\sigma=0.5,1,2$. The higher curves correspond to the wider
$\sigma$'s. All quantities are given in $0.1 p_0$
units.\label{v2grafdecophoton3}}
\end{figure}

We plot in Figure \ref{v2grafdecophoton1} the negativity $N$ in Eq.
(\ref{v2deco11ph}) as a function of $\tilde{B}L$ for $p_0=10$,
$\sigma=2$, and $w_0=10$. All quantities are given in $0.1 p_0$
units. The entanglement decreases as the magnetic field $\tilde{B}$
or the length $L$ of the dielectric increase. We plot also in Figure
\ref{v2grafdecophoton2} the negativity $N$ as a function of $\sigma$
for $p_0=10$, $\tilde{B}L=4$, and $w_0=10$. Surprisingly, and
opposite to the two-fermion case, the entanglement increases as the
momentum width $\sigma$ of the photon is larger. This effect comes
from the fact that for larger widths, centered at $w_0$, the
contribution from the region around the resonance frequency $w_0$,
in which the effect of the medium is appreciable, becomes smaller.
On the other hand, in the limit of negligible width, the spin could
not get entangled with the momentum so in this limit the spin-spin
entanglement would remain maximal. We observe then that there is a
region of intermediate widths $\sigma$ in which the spin-spin
entanglement becomes minimal. Finally, we plot in Figure
\ref{v2grafdecophoton3} the negativity $N$ as a function of $w_0$
for $p_0=10$, $\tilde{B}L=2$, and $\sigma=0.5,1,2$. The higher
curves correspond to the wider $\sigma$'s. These graphics show that
the entanglement decreases mainly for resonance frequencies $w_0$
around the average momentum $p_0$. It also shows the surprising
behavior mentioned above: For wider $\sigma$, the entanglement is
larger, and the interval of $w_0$ for which the entanglement
decreases is wider, as expected according to the previous analysis.

\section{IS PURELY QUANTUM COMMUNICATION FEASIBLE?\label{lbsetc4}}

A cautious reader may immediately object that, in principle, our
preceding analysis seems to suggest the feasibility of communication
through a purely quantum channel, i.e. without classical
communication, contrary to all known quantum-informational
protocols. Let us illustrate this point with the following proposal
based in the previous two-fermion case. As usual, Alice and Bob will
be the corresponding observers of each fermion. The joint spin-spin
state is given by equation \eqref{v2deco6}, which immediately drives
one to the subsequent reduced spin state for fermion $B$
\begin{equation}
\rho_{B}=\begin{pmatrix}
I_{\downarrow\downarrow} & 0\\
0 & I_{\uparrow\uparrow}
\end{pmatrix}.
\end{equation}
But the quantities $I_{ss'}$ depend on the magnetic field $B_{0}$
(cf.\ equations \eqref{v2deco7} and \eqref{TransCoef}), \emph{which
can be controlled by Alice}. This allows them to agree on the
following procedure. They agree on communicating with a binary
alphabet with classical bits $0$ and $1$. If Alice were to
communicate $0$, she would adjust $B_{0}$ so that the reduced spin
state for Bob is, for example,
\begin{equation}
\rho_{B}^{(0)}=\begin{pmatrix}
\frac{3}{4} & 0\\
0 & \frac{1}{4}
\end{pmatrix}.
\end{equation}
They prepare a statistically significative amount of pairs of
fermions under such conditions. Then Bob, when measuring the spin
upon his fermion, will typically obtain spin up in the $75\%$ of the
measurements and spin down in the remaining $25\%$. He thus deduces
that Alice is sending the bit $0$.

 On the contrary, if Alice were to
communicate the bit $1$, she would adjust $B_{0}$ so that the
reduced spin state for Bob is, for example,
\begin{equation}
\rho_{B}^{(1)}=\begin{pmatrix}
\frac{1}{4} & 0\\
0 & \frac{3}{4}
\end{pmatrix}.
\end{equation}
They also prepare a statistically significative amount of pairs and
Bob performs his measurements. He will detect $75\%$ of them in the
spin down state and $25\%$ in the spin up state. He thus deduces
that Alice is sending the bit $1$. Notice that this information
transmission is carried out without the assistance of classical
communication.

The flaw stems from the disregarding of the wave packet reflection
in Alice's site. This can be seen in two complementary ways. On the
one hand, to perform a genuine information transmission in which
Bob's fermion is actually carrying information encoded by Alice, he
must be able to discern between those fermions whose pairs have been
reflected in Alice's barrier potential, so that he can securely
discard them (they are not carrying information at all). And this is
only possible if Alice \emph{classically} communicates this
information to Bob. On the other hand, a detailed calculation taking
into account the reflection coefficients, hence without the
normalization appearing in the denominator of \eqref{v2deco7}, shows
that Bob's reduced spin state will be given by
\begin{eqnarray}
\rho_{B}&=&\frac{1}{2}\begin{pmatrix}\int d^{3}\mathbf{p}|\mathcal{G}(\mathbf{p})|^{2}|\mathcal{R}_{\downarrow}(p)|^{2} & 0\\0 &\int d^{3}\mathbf{p}|\mathcal{G}(\mathbf{p})|^{2}|\mathcal{R}_{\uparrow}(p)|^{2}\end{pmatrix}\nonumber\\
&+&\frac{1}{2}\begin{pmatrix}\int d^{3}\mathbf{p}|\mathcal{G}(\mathbf{p})|^{2}|\mathcal{T}_{\downarrow}(p)|^{2} & 0\\0 &\int d^{3}\mathbf{p}|\mathcal{G}(\mathbf{p})|^{2}|\mathcal{T}_{\uparrow}(p)|^{2}\end{pmatrix}=\nonumber\\
&=&\frac{1}{2}\begin{pmatrix}1&0\\
0&1\end{pmatrix},
\end{eqnarray}
where $\mathcal{R}_{s}(p)$ denotes the corresponding reflection
coefficient for spin $s$ and momentum $p$. The calculation reveals
that Bob gains no information whatsoever from Alice's decisions,
unless she classically informs Bob about them.

Mathematically this can be expressed through the unitary character
of the process. If no classical information is exchanged, the
evolution is locally unitary ($\Psi^{(a)}\otimes\Psi^{(b)}\to
U^{(a)}[\Psi^{(a)}]\otimes U^{(b)}[\Psi^{(b)}]$) and thus cannot
change the entanglement shared by both parties (the entanglement is
invariant under local unitary evolution). Consequently no
information through the purely quantum channel can be obtained. On
the contrary, if Alice classically sends information to Bob, she is
actually selecting a subset of her incoming fermions, i.e.\ she is
\emph{projecting} her state ($\Psi^{(a)}\otimes\Psi^{(b)}\to
P^{(ab)}[\Psi^{(a)}\otimes\Psi^{(b)}]$, where $P^{(ab)}$ is an
orthogonal projector\footnote{More generally, it can also be a POVM,
depending on whether the information provided by Alice is complete
or not \cite{P95}.}), which is a non-unitary operator which changes
the entanglement class. This fact allows them to exploit the initial
quantum correlations between their fermions to establish a
communication protocol.

In summary, this example shows once more the impossibility of using
quantum correlations, i.e.\ entanglement to exchange information
without the aid of classical communication.

\section{CONCLUSIONS\label{lbsetc5}}
We showed the spin entanglement loss by transfer of correlations to
the momentum of one of the particles, through a local spin-momentum
entangling interaction. This phenomenon, already analyzed for a
non-interacting particular case in the context of Wigner rotations
of special relativity, may produce decoherence of Bell spin states.
The momentum of each particle is a very simple reservoir and indeed
it is one that cannot be eliminated by improving the experimental
conditions, due to Heisenberg's principle. We show that an
$s=\frac{1}{2}$ fermion (photon), which initially belongs to a Bell
spin state, may lose its spin correlations due to this physical
phenomenon when traversing a local magnetic field (optically-active
medium). These specific media entangle each component of the spin
state of the particle with its momentum, like in a Stern-Gerlach
device. This could have implications for quantum communication and
information processing devices.
\section*{ACKNOWLEDGMENTS}
This work was partially supported by the Spanish MEC projects No.
FIS2005-05304 (L.L. and J.L.) and No. FIS2004-01576 (D.S.). L.L.
acknowledges support from the FPU grant No. AP2003-0014.


\begin{thebibliography}{17}
\expandafter\ifx\csname
natexlab\endcsname\relax\def\natexlab#1{#1}\fi
\expandafter\ifx\csname bibnamefont\endcsname\relax
  \def\bibnamefont#1{#1}\fi
\expandafter\ifx\csname bibfnamefont\endcsname\relax
  \def\bibfnamefont#1{#1}\fi
\expandafter\ifx\csname citenamefont\endcsname\relax
  \def\citenamefont#1{#1}\fi
\expandafter\ifx\csname url\endcsname\relax
  \def\url#1{\texttt{#1}}\fi
\expandafter\ifx\csname urlprefix\endcsname\relax\def\urlprefix{URL
}\fi \providecommand{\bibinfo}[2]{#2}
\providecommand{\eprint}[2][]{\url{#2}}

\bibitem[{\citenamefont{Nielsen and Chuang}(2000)}]{nielsen}
\bibinfo{author}{\bibfnamefont{M.~A.} \bibnamefont{Nielsen}} \bibnamefont{and}
  \bibinfo{author}{\bibfnamefont{I.~L.} \bibnamefont{Chuang}},
  \emph{\bibinfo{title}{Quantum Computation and Quantum Information}}
  (\bibinfo{publisher}{Cambridge University Press, Cambridge},
  \bibinfo{year}{2000}).

\bibitem[{\citenamefont{Galindo and Mart\'{\i}n-Delgado}(2002)}]{GM02}
\bibinfo{author}{\bibfnamefont{A.}~\bibnamefont{Galindo}} \bibnamefont{and}
  \bibinfo{author}{\bibfnamefont{M.~A.} \bibnamefont{Mart\'{\i}n-Delgado}},
  \bibinfo{journal}{Rev. Mod. Phys.} \textbf{\bibinfo{volume}{74}},
  \bibinfo{pages}{347} (\bibinfo{year}{2002}).


\bibitem[{\citenamefont{Yu and Eberly}(2002)}]{Ye02}
\bibinfo{author}{\bibfnamefont{T.}~\bibnamefont{Yu}} \bibnamefont{and}
  \bibinfo{author}{\bibfnamefont{J.~H.} \bibnamefont{Eberly}},
  \bibinfo{journal}{Phys. Rev. B} \textbf{\bibinfo{volume}{66}},
  \bibinfo{pages}{193306} (\bibinfo{year}{2002}).

\bibitem[{\citenamefont{Yu and Eberly}(2003)}]{Ye03}
\bibinfo{author}{\bibfnamefont{T.}~\bibnamefont{Yu}} \bibnamefont{and}
  \bibinfo{author}{\bibfnamefont{J.~H.} \bibnamefont{Eberly}},
  \bibinfo{journal}{Phys. Rev. B} \textbf{\bibinfo{volume}{68}},
  \bibinfo{pages}{165322} (\bibinfo{year}{2003}).

\bibitem[{\citenamefont{Yu and Eberly}(2004)}]{Ye04}
\bibinfo{author}{\bibfnamefont{T.}~\bibnamefont{Yu}} \bibnamefont{and}
  \bibinfo{author}{\bibfnamefont{J.~H.} \bibnamefont{Eberly}},
  \bibinfo{journal}{Phys. Rev. Lett.} \textbf{\bibinfo{volume}{93}},
  \bibinfo{pages}{140404} (\bibinfo{year}{2004}).

\bibitem[{\citenamefont{Dodd and Halliwell}(2004)}]{DH04}
\bibinfo{author}{\bibfnamefont{P.~J.} \bibnamefont{Dodd}}
  \bibnamefont{and} \bibinfo{author}{\bibfnamefont{J.~J.}~\bibnamefont{Halliwell}},
  \bibinfo{journal}{Phys. Rev. A} \textbf{\bibinfo{volume}{69}},
  \bibinfo{pages}{052105} (\bibinfo{year}{2004}).

\bibitem[{\citenamefont{Nikoli\'c and Souma}(2005)}]{NS05}
\bibinfo{author}{\bibfnamefont{B.~K.} \bibnamefont{Nikoli\'c}}
  \bibnamefont{and} \bibinfo{author}{\bibfnamefont{S.}~\bibnamefont{Souma}},
  \bibinfo{journal}{Phys. Rev. B} \textbf{\bibinfo{volume}{71}},
  \bibinfo{pages}{195328} (\bibinfo{year}{2005}).

\bibitem[{\citenamefont{Carvalho et~al.}(2005)\citenamefont{Carvalho, Mintert,
  Palzer, and Buchleitner}}]{CMPB05}
\bibinfo{author}{\bibfnamefont{A.~R.~R.} \bibnamefont{Carvalho}},
  \bibinfo{author}{\bibfnamefont{F.}~\bibnamefont{Mintert}},
  \bibinfo{author}{\bibfnamefont{S.}~\bibnamefont{Palzer}}, \bibnamefont{and}
  \bibinfo{author}{\bibfnamefont{A.}~\bibnamefont{Buchleitner}},
  \bibinfo{journal}{quant-ph/0508114}  (\bibinfo{year}{2005}).

\bibitem[{\citenamefont{Fran\c{c}a Santos et~al. et~al.}(2005)\citenamefont{Fran\c{c}a Santos, Milman,
  Davidovich, and Zagury}}]{FMDZ05}
\bibinfo{author}{\bibfnamefont{M.} \bibnamefont{Fran\c{c}a Santos}},
  \bibinfo{author}{\bibfnamefont{P.}~\bibnamefont{Milman}},
  \bibinfo{author}{\bibfnamefont{L.}~\bibnamefont{Davidovich}}, \bibnamefont{and}
  \bibinfo{author}{\bibfnamefont{N.}~\bibnamefont{Zagury}},
  \bibinfo{journal}{Phys. Rev. A}   \textbf{\bibinfo{volume}{73}},
  \bibinfo{pages}{040305(R)} (\bibinfo{year}{2006}).


\bibitem[{\citenamefont{Peres et~al.}(2002)\citenamefont{Peres, Scudo, and
  Terno}}]{PST02}
\bibinfo{author}{\bibfnamefont{A.}~\bibnamefont{Peres}},
  \bibinfo{author}{\bibfnamefont{P.~F.} \bibnamefont{Scudo}}, \bibnamefont{and}
  \bibinfo{author}{\bibfnamefont{D.~R.} \bibnamefont{Terno}},
  \bibinfo{journal}{Phys. Rev. Lett.} \textbf{\bibinfo{volume}{88}},
  \bibinfo{pages}{230402} (\bibinfo{year}{2002}).

\bibitem[{\citenamefont{Czachor}(2005)}]{C05}
\bibinfo{author}{\bibfnamefont{M.}~\bibnamefont{Czachor}},
  \bibinfo{journal}{Phys. Rev. Lett.} \textbf{\bibinfo{volume}{94}},
  \bibinfo{pages}{078901} (\bibinfo{year}{2005}).

\bibitem[{\citenamefont{Alsing and Milburn}(2002)}]{AM02}
\bibinfo{author}{\bibfnamefont{P.~M.} \bibnamefont{Alsing}} \bibnamefont{and}
  \bibinfo{author}{\bibfnamefont{G.~J.} \bibnamefont{Milburn}},
  \bibinfo{journal}{Quant. Inf. Comput.} \textbf{\bibinfo{volume}{2}},
  \bibinfo{pages}{487} (\bibinfo{year}{2002}).

\bibitem[{\citenamefont{Gingrich and Adami}(2002)}]{GA02}
\bibinfo{author}{\bibfnamefont{R.~M.} \bibnamefont{Gingrich}} \bibnamefont{and}
  \bibinfo{author}{\bibfnamefont{C.}~\bibnamefont{Adami}},
  \bibinfo{journal}{Phys. Rev. Lett.} \textbf{\bibinfo{volume}{89}},
  \bibinfo{pages}{270402} (\bibinfo{year}{2002}).

\bibitem[{\citenamefont{Pachos and Solano}(2003)}]{PS03}
\bibinfo{author}{\bibfnamefont{J.}~\bibnamefont{Pachos}} \bibnamefont{and}
  \bibinfo{author}{\bibfnamefont{E.}~\bibnamefont{Solano}},
  \bibinfo{journal}{Quant. Inf. Comput.} \textbf{\bibinfo{volume}{3}},
  \bibinfo{pages}{115} (\bibinfo{year}{2003}).

\bibitem[{\citenamefont{Czachor and Wilczewski}(2003)}]{C03}
\bibinfo{author}{\bibfnamefont{M.}~\bibnamefont{Czachor}} \bibnamefont{and}
  \bibinfo{author}{\bibfnamefont{M.}~\bibnamefont{Wilczewski}},
  \bibinfo{journal}{Phys. Rev. A} \textbf{\bibinfo{volume}{68}},
  \bibinfo{pages}{010302(R)} (\bibinfo{year}{2003}).

\bibitem[{\citenamefont{Gingrich et~al.}(2003)\citenamefont{Gingrich, Bergou,
  and Adami}}]{GBA03}
\bibinfo{author}{\bibfnamefont{R.~M.} \bibnamefont{Gingrich}},
  \bibinfo{author}{\bibfnamefont{A.~J.} \bibnamefont{Bergou}},
  \bibnamefont{and} \bibinfo{author}{\bibfnamefont{C.}~\bibnamefont{Adami}},
  \bibinfo{journal}{Phys. Rev. A} \textbf{\bibinfo{volume}{68}},
  \bibinfo{pages}{042102} (\bibinfo{year}{2003}).

\bibitem[{\citenamefont{Lamata et~al.}(2005)\citenamefont{Lamata,
  Mart\'{\i}n-Delgado, and Solano}}]{LMDS05}
\bibinfo{author}{\bibfnamefont{L.}~\bibnamefont{Lamata}},
  \bibinfo{author}{\bibfnamefont{M.~A.} \bibnamefont{Mart\'{\i}n-Delgado}},
  \bibnamefont{and} \bibinfo{author}{\bibfnamefont{E.}~\bibnamefont{Solano}},
  \bibinfo{journal}{quant-ph/0512081}  (\bibinfo{year}{2005}).

\bibitem[{\citenamefont{Scully et al.}(2005)\citenamefont{Scully and Zubairy}}]{SZ97}
\bibinfo{author}{\bibfnamefont{M.~O.}~\bibnamefont{Scully}},  \bibnamefont{and}
  \bibinfo{author}{\bibfnamefont{M.~S.} \bibnamefont{Zubairy}},
   \emph{\bibinfo{title}{Quantum Optics}}
  (\bibinfo{publisher}{Cambridge University Press, Cambridge},
  \bibinfo{year}{1997}).

\bibitem[{\citenamefont{Fran\c{c}a Santos et~al.}(2005)\citenamefont{Fran\c{c}a Santos,
  Milman, Khoury, and Souto Ribeiro}}]{FMKS01}
\bibinfo{author}{\bibfnamefont{M.}~\bibnamefont{Fran\c{c}a Santos}},
  \bibinfo{author}{\bibfnamefont{P.} \bibnamefont{Milman}},
  \bibinfo{author}{\bibfnamefont{A.~Z.}~\bibnamefont{Khoury}},
  \bibnamefont{and} \bibinfo{author}{\bibfnamefont{P.~H.}~\bibnamefont{Souto Ribeiro}},
  \bibinfo{journal}{Phys. Rev. A} \textbf{\bibinfo{volume}{64}},
  \bibinfo{pages}{023804} (\bibinfo{year}{2001}).


\bibitem[{\citenamefont{Vidal and Werner}(2002)}]{VW02}
\bibinfo{author}{\bibfnamefont{G.}~\bibnamefont{Vidal}} \bibnamefont{and}
  \bibinfo{author}{\bibfnamefont{R.~F.} \bibnamefont{Werner}},
  \bibinfo{journal}{Phys. Rev. A} \textbf{\bibinfo{volume}{65}},
  \bibinfo{pages}{032314} (\bibinfo{year}{2002}).

\bibitem[{\citenamefont{Peres}(1995)}]{P95}
\bibinfo{author}{\bibfnamefont{A.}~\bibnamefont{Peres}},
  \emph{\bibinfo{title}{Quantum Theory: Concepts and Methods}}
  (\bibinfo{publisher}{Kluwer Academic Publishers, Dordrecht},
  \bibinfo{year}{1995}).

\end{thebibliography}
\end{document}